\documentclass[prd,aps]{revtex4}
\usepackage{amssymb,amsmath,amsfonts,amsbsy,epsfig}
\usepackage{slashed}

\newcommand{\be}{\begin{equation}}
\newcommand{\ee}{\end{equation}}
\newcommand{\bea}{\begin{eqnarray}}
\newcommand{\eea}{\end{eqnarray}}
\newcommand{\nn}{\nonumber}

\def\L{{\mathcal L}}
\def\M{{\mathcal M}}

\begin{document}

\title{On the photon mass in Very Special Relativity}
\date{\today}

\author{
Jorge Alfaro$^{a}$, and Alex Soto$^{a}$
}

\affiliation{
$^{a}$Instituto de F\'{i}sica, Pontificia Universidad de Cat\'olica de Chile, \mbox{Av. Vicu\~na Mackenna 4860, Santiago, Chile}
}

\begin{abstract} 

In Very Special Relativity (VSR) a gauge invariant photon mass is allowed. We explore some consequences of keeping this parameter not zero. VSR-Maxwell equations are presented. In addition, we show modifications to the Feynman rules keeping the photon mass and we apply it in the computation of the electron self energy, whose result is free from infrared divergences. A computation of the Coulomb scattering is developed and a small signal of the privileged direction predicted in the theory is present at tree level. Radiative corrections are computed and the photon mass cancels due to soft photon emission as in the standard case. 

\end{abstract}

\maketitle 

\section{Introduction}

The absence of any new particle in colliders allows us the possibility to explore new theories to give an answer to  unsolved problems in the standard model of particles. Very Special Relativity (VSR)\cite{Cohen:2006ky} is a proposal where without new particles, it is possible to explain the neutrino mass, considering the universe invariant under a subgroup of the full Lorentz group. In the reference \cite{Cohen:2006ir}  the SIM(2) group is used to explain how new terms, invariant under this group, allow the existence of neutrino mass preserving the same kinematics than special relativity because in this group there are not invariant tensors, besides the ones which are invariant under the Lorentz group. As a consequence of this model, non local terms appear and a tiny Lorentz violation parametrized by the neutrino mass is expected.\\

 In Poincare invariant electrodynamics (QED) the photon mass is strictly zero because it is forbidden by gauge invariance. Gauge invariance is crucial to prove that unitarity and renormalizability are  satisfied in  QED\cite{Peskin:1995ev}. However, in VSR the possibility of a gauge invariant non-zero photon mass appears\cite{Cheon:2009zx, Alfaro:2013uva, Alfaro:2015fha}. The introduction of the mass terms for neutrino and photon in VSR theories do not affect the renormalizability of the model, because the additional SIM(2)-invariant expressions contain the non-local term $1/n\cdot\partial$, which in the large momentum limit vanishes, leaving the same ultraviolet behavior as in the Lorentz invariant theories. The model is renormalizable and unitary (some comments on this can be found in \cite{Alfaro:2013uva,Alfaro:2015fha}).\\

Experimental tests have shown the photon mass is almost zero or at least too small, for example, Bonetti et. al\cite{Bonetti:2016cpo} have reported an upper bound of $1.8\cdot10^{-14}$ eV using radio bursts, while the Particle Data Group\cite{Tanabashi:2018oca} in 2018 have published a bound of $10^{-18}$ eV.  New tests with higher precision should be carried on in the future. Although the mass is considered to be zero, the possibility it should be different from zero without breaking gauge invariance in VSR is open. The aim of this work is to  present an analysis of some consequences if we let the photon be massive within the theory, which is unitary, renormalizable and gives an answer to the neutrino mass.\\

The paper is organized as follows: in section II we present the VSR formalism in the QED sector considering the photon mass. In section III we derive the Feynman rules associated to the lagrangian of QED-VSR and some relations between the vertex in the theory useful to construct the Ward identities. In section IV we compute the electron self energy and the Ward identity. In section V we compute the cross section at tree level of the Coulomb Scattering. The section VI is devoted to show the cancellation of the infrared divergencies in the one loop radiative corrections. Finally in section VII we present the conclusions and some remarks.

\section{QED VSR Model with photon mass}
In this section we will review the references \cite{Alfaro:2013uva, Alfaro:2015fha} to fix the frame for the next calculations. We start with a gauge field $A_\mu$ and a scalar field $\phi$ coupled with this gauge field,  whose the gauge transformations are
\bea
\delta A_\mu=\tilde{\partial}_\mu \Lambda,\\
\delta\phi=i\Lambda\phi,
\eea
where we have defined the wiggle operator
\be
\label{wiggle}
\tilde{\partial}_\mu=\partial_\mu+\frac{1}{2}\frac{m_\gamma^2}{n\cdot \partial}n_\mu,
\ee
and $m_\gamma$ is a parameter with mass dimension to fix the units, $n_\mu$ is the constant null vector which is present in VSR theories. We can define the covariant derivative
\be
D_\mu\phi=\partial_\mu\phi-i A_\mu\phi-\frac{i}{2}m_\gamma^2n_\mu\left(\frac{1}{(n\cdot \partial)^2}n\cdot A\right)\phi.
\ee
We can compute the field strength related to $D_\mu$ as
\be
[D_\mu,D_\nu]\phi=-i F_{\mu\nu}\phi.
\ee
Thus, we get
\be
F_{\mu\nu}=\partial_\mu A_\nu-\partial_\nu A_\mu+\frac{1}{2}m_\gamma^2n_\nu\left(\frac{1}{(n\cdot \partial)^2}\partial_\mu(n\cdot A)\right)-\frac{1}{2}m_\gamma^2n_\mu\left(\frac{1}{(n\cdot \partial)^2}\partial_\nu(n\cdot A)\right).
\ee
We notice that a redefinition of $A_\mu\rightarrow A_\mu-\frac{1}{2}m_\gamma^2n_\mu((n\cdot \partial)^{-2}(n\cdot A))$ does not affect the observables and eliminates the modification by the mass factor. So, after this redefinition, we get the same electromagnetic tensor given by $F_{\mu\nu}=\partial_\mu A_\nu-\partial_\nu A_\mu$ and the same gauge transformation as in the Electroweak Standard Model given by
\be
\label{gt}
A_\mu \to A_\mu+\partial_\mu \Lambda
\ee

However, we can define a new VSR field strenght  $\tilde{F}_{\mu\nu}$ using the wiggle derivative in the way $\tilde{F}_{\mu\nu}=\tilde{\partial}_\mu A_\nu-\tilde{\partial}_\nu A_\mu$, and we can write it in terms only of the standard $F_{\mu\nu}$ as
\be
\tilde{F}_{\mu\nu}=F_{\mu\nu}+\frac{1}{2}m_\gamma^2\left[n_\nu\frac{1}{(n\cdot\partial)^2}(n^\alpha F_{\mu\alpha})-n_\mu\frac{1}{(n\cdot\partial)^2}(n^\alpha F_{\nu\alpha})\right].
\ee


 We notice that $\tilde{F}_{\mu\nu}$,  is not Lorentz but SIM(2) invariant.  From its definition, we see that $\tilde{F}_{\mu\nu}$ is invariant under equation (\ref{gt}) too. This fact is important because the lagrangian that we will construct below can be written in terms of $\tilde{F}_{\mu\nu}$ and the parameter $m_\gamma$ will be identified with the photon mass(please see equation (\ref{eomm})). With this we will have a theory with a photon mass coming from a term invariant under the standard gauge transformations of $A_\mu$ in  equation (\ref{gt}) .\\


We can use the VSR field strength expression to write the VSR gauge lagrangian as
\be
\L_{gauge}=-\frac{1}{4}\tilde{F}_{\mu\nu}\tilde{F}^{\mu\nu}.
\ee
Therefore, the lagrangian is
\be
\label{lagrangian}
\L_{gauge}=-\frac{1}{4}F_{\mu\nu}F^{\mu\nu}-\frac{1}{2}m_\gamma^2(n^\alpha F_{\mu\alpha})\frac{1}{(n\cdot\partial)^2}(n_\beta F^{\mu\beta}).
\ee

From equation (\ref{lagrangian}) we get the equation of motion for $F_{\mu\nu}$ with the help of the wiggle derivative,
\be
\label{eom}
\partial_\mu F^{\mu\nu}+\frac{1}{2}m_\gamma^2n^\nu\frac{1}{(n\cdot\partial)^2}\partial_\alpha(n_\beta F^{\alpha\beta})+m_\gamma^2\frac{1}{n\cdot\partial}(n_\beta F^{\beta\nu})=0.
\ee
We write equation (\ref{eom}) in terms of the gauge field as
\be
\label{eoma}
\partial^2A^\nu-\partial^\nu\partial_\mu A^\mu+\frac{1}{2}m_\gamma^2n^\nu\frac{1}{(n\cdot\partial)^2}[\partial^2(n\cdot A)-(n\cdot\partial)(\partial_\alpha A^\alpha)]+m_\gamma^2\frac{1}{n\cdot\partial}(n\cdot\partial A^\nu-\partial^\nu n\cdot A)=0.
\ee
If we contract equation (\ref{eoma}) with $n_\nu$ we get
\be
\partial^2n\cdot A-n\cdot\partial(\partial\cdot A)=0.\label{eq1}
\ee

We will fix the Lorentz gauge $\partial_\mu A^\mu=0$. Then equation (\ref{eq1}) implies the condition
\be
\label{na}
\partial^2(n\cdot A)=0.
\ee
However, a gauge degree of freedom remains, since $\partial^\mu A'_\mu=\partial^\mu A_\mu+\partial^2\Lambda_1$ implies $\partial^2\Lambda_1=0$, if both $A'_\mu$ and $A_\mu$ are in  the Lorentz gauge. We will use the remaining gauge freedom to impose the additional condition $n\cdot A=0$. In fact:
\be
n\cdot A'=n\cdot A+n\cdot\partial\Lambda_1=0
\ee
has the solution $\Lambda_1=-\frac{1}{n\cdot\partial}n\cdot A$. To complete our proof, we have to verify that $\partial^2\Lambda_1=0$. Thus
\be
\partial^2\Lambda_1=-\frac{1}{n\cdot\partial}\partial^2 n\cdot A=0
\ee
In the last step we used equation(\ref{na}).

Therefore, we apply the Lorentz gauge plus the subsidiary condition $n\cdot A=0$ in the equation (\ref{eoma}) and we obtain
\be
\label{eomm}
(\partial^2+m_\gamma^2)A^\nu=0.
\ee
We see from equation (\ref{eomm}) that $A^\nu$ is a field with mass $m_\gamma$. Hence, in VSR we have a photon mass coming from a term that is gauge invariant, unlike the case with a term of the type $m_\gamma^2A^\mu A_\mu$, which is not.\\

Now, we use a plane wave solution for $A^\nu$, as $A^\nu=\varepsilon^\nu e^{-ikx}$, so we have
\be
k^2-m_\gamma^2=0
\ee
with the conditions $k\cdot\varepsilon=0$ and $n\cdot\varepsilon=0$. So, the gauge field has two polarizations, as in the standard case.\\

\section{VSR Maxwell Equations}
We can use the equation of motion in (\ref{eom}) to get the modified Gauss law and the Ampere law in the VSR framework putting $\mu=0$ and $\mu=j$, where $j=1,2,3$. So, using the expressions for $\vec{E}$ and $\vec{B}$ these two first Maxwell equations in vacuum are
\bea
\vec{\nabla}\cdot \vec{E}+\frac{1}{2}m_\gamma^2\frac{n^0}{(n\cdot\partial)^2}\left[n^0\vec{\nabla}\cdot \vec{E}+\vec{n}\cdot\left(\vec{\nabla}\times \vec{B}-\frac{\partial \vec{E}}{\partial t}\right)\right]+m_\gamma^2\frac{1}{n\cdot\partial}(\vec{n}\cdot\vec{E})=0,\\
\vec{\nabla}\times \vec{B}-\frac{\partial\vec{E}}{\partial t}+\frac{1}{2}m_\gamma^2\frac{\vec{n}}{(n\cdot\partial)^2}\left[n^0\vec{\nabla}\cdot \vec{E}+\vec{n}\cdot\left(\vec{\nabla}\times \vec{B}-\frac{\partial \vec{E}}{\partial t}\right)\right]+m_\gamma^2\frac{1}{n\cdot\partial}(\vec{n}\times\vec{B}-n^0\vec{E})=0.
\eea
 We use the Bianchi identity $\tilde{\partial}_{\left[\mu\right.}\tilde{F}_{\left.\nu\rho\right]}=0$, to get the last two Maxwell equations

\bea
\left(1+\frac{1}{2}m_\gamma^2\frac{\vert \vec{n}\vert^2}{(n\cdot \partial)^2}\right)\frac{\partial\vec{B}}{\partial t}+\left(1+\frac{1}{2}m_\gamma^2\frac{(n^0)^2}{(n\cdot \partial)^2}\right)\vec{\nabla}\times\vec{E}+\frac{1}{2}m_\gamma^2\frac{1}{n\cdot\partial}(n^0\vec{B}+\vec{n}\times\vec{E})\nn\\
-\frac{1}{2}m_\gamma^2\frac{1}{(n\cdot\partial)^2}\left[n^0(\vec{n}\cdot\vec{\nabla})\vec{B}-n^0\vec{n}(\vec{\nabla}\cdot\vec{B})+\vec{n}\times\vec{\nabla}(\vec{n}\cdot\vec{E})+n^0\vec{n}\times\frac{\partial\vec{E}}{\partial t}+\vec{n}\frac{\partial(\vec{n}\cdot\vec{B})}{\partial t}\right]&=&0\\
\left(1+\frac{1}{2}m_\gamma^2\frac{\vert\vec{n}\vert^2}{(n\cdot\partial)^2}\right)\vec{\nabla}\cdot\vec{B}+\frac{1}{2}m_\gamma^2\frac{1}{n\cdot\partial}\vec{n}\cdot\vec{B}+\frac{1}{2}m_\gamma^2\frac{1}{(n\cdot\partial)^2}\left[n^0\vec{n}\cdot(\vec{\nabla}\times\vec{E})-\vec{n}\cdot\vec{\nabla}(\vec{n}\cdot\vec{B})\right]&=&0
\eea

We can see that setting $m_\gamma^2=0$, we recover the standard Maxwell equations. We remark the VSR photon mass parameter is very small to be significative in a classical experiment, so, we will not focus on this line.

\section{Feynman Rules}
We will review in brief the Feynman rules for the model with photon mass. In order to get them, we write the gauge lagrangian in equation (\ref{lagrangian}) plus a gauge fixing term as follows
\be
\L_{gauge+gf}=-\frac{1}{4}F_{\mu\nu}F^{\mu\nu}-\frac{1}{2}m_\gamma^2(n^\alpha F_{\mu\alpha})\frac{1}{(n\cdot\partial)^2}(n_\beta F^{\mu\beta})-\frac{1}{2\xi}(\partial_\mu A^\mu)^2.
\ee
Performing some algebraic manipulations we get
\be
\L_{gauge+gf}=\frac{1}{2}A_\nu\left[(\partial^2+m_\gamma^2)g^{\mu\nu}-\left(1-\frac{1}{\xi}\right)\partial^\mu\partial^\nu+m_\gamma^2\frac{n^\mu n^\nu\partial^2}{(n\cdot\partial)^2}-m_\gamma^2\frac{\partial^\mu n^\nu+\partial^\nu n^\mu}{n\cdot\partial}\right]A_\mu.
\ee
Then, in Fourier space and with a little bit of algebra we have an expression for the propagator and we choose the Feynman gauge $\xi=1$, so
\be
\label{propph}
\Delta(p,\bar{n})=-\frac{i}{p^2-m_\gamma^2}\left[g_{\mu\nu}+\frac{m_\gamma^2}{(n\cdot p)_{\bar{n}}^2}n_\mu n_\nu-\frac{m_\gamma^2}{p^2(n\cdot p)_{\bar{n}}}(p_\mu n_\nu+p_\nu n_\mu)\right],
\ee
where here we have assigned the subindex $\bar{n}$ for $n\cdot p$ in the denominators, related with the Mandelstam-Leibbrandt prescription\cite{Mandelstam:1982cb,Leibbrandt:1983pj}

\be
\label{M-L}
\frac{1}{n\cdot p}=\lim_{\epsilon\to0}\frac{\bar{n}\cdot p}{(n\cdot p)(\bar{n}\cdot p)+i\epsilon}.
\ee

It is important to notice, if we set $m_\gamma=0$ in equation (\ref{propph}) we recover the standard propagator in the Feynman gauge.\\

If now we allow interaction with an electron, the new lagrangian will be
\be
\L=\L_{gauge+gf}+\L_{fermion},
\ee
where
\be
\label{lferm}
\L_{fermion}=\bar{\psi}\left(i\slashed{D}-M+i\frac{1}{2}m^2\slashed{n}\frac{1}{n\cdot D}\right)\psi,
\ee
and here $m$ is the neutrino mass, $M$ the acquired mass of the electron after spontaneous symmetry breaking, and $D_\mu=\partial_\mu+ieA_\mu$.\\

\begin{figure}[h!]
\centering
\includegraphics[scale=0.08]{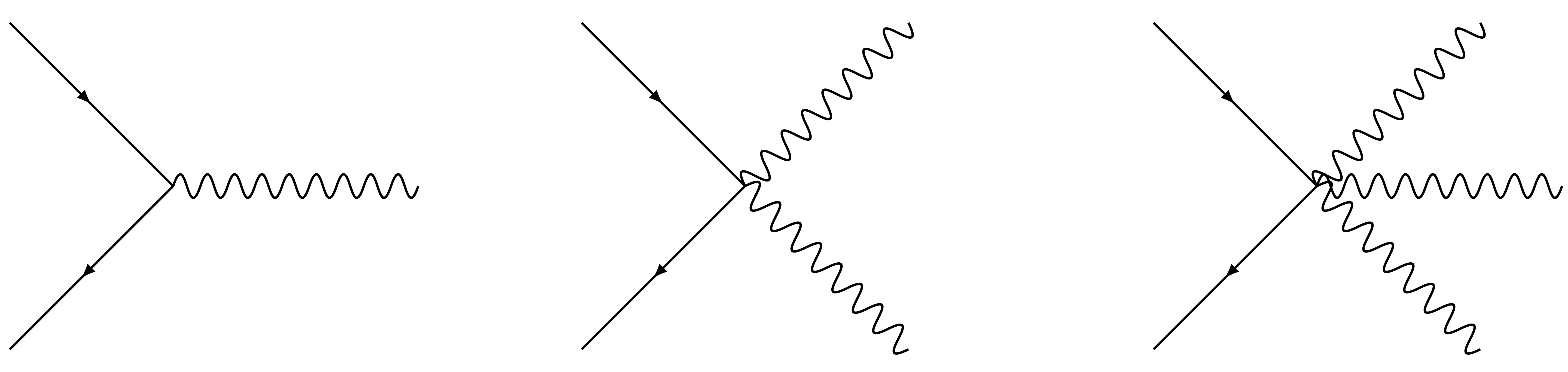}
\caption{Examples of vertices in VSR. They come from the perturbative expansion of the non-local term $(n\cdot D)^{-1}$.}
\label{fig:vertex}
\end{figure}

From equation (\ref{lferm}) we can get the electron propagator
\be
S_F(p,\bar{n})=i\frac{\slashed{p}+M-\frac{m^2}{2}\frac{\slashed{n}}{(n\cdot p)_{\bar{n}}}}{p^2-M_e^2+i\varepsilon},
\ee
where $M_e^2=M^2+m^2$. In addition, because the nonlocality of the $(n\cdot D)^{-1}$ we can have new kind of vertex, with more external photon legs (see figure \ref{fig:vertex}).\\

We can establish a relationship between the vertex with a number of external photon legs and the vertex with one less external photon legs. We start with the standard vertex as is shown in the left diagram in figure \ref{fig:vertex}. The mathematical expression is easy to get from equation (\ref{lferm}). Then,
\be
\label{vert}
V_{1\mu}[(p,\bar{n}_1),(p+q,\bar{n}_2)]=\gamma_\mu+\frac{1}{2}m^2\slashed{n}\frac{n_\mu}{(n\cdot p)_{\bar{n}_1}[n\cdot(p+q)]_{\bar{n}_2}},
\ee

where here we have two differents $\bar{n}$ for each momentum in the denominator with $n\cdot p$. Our choice is general, nevertheless the Ward Identity will determine the relationships as we will see below.\\ 

If we contract the vertex with the photon external momentum $q$ in the equation (\ref{vert}) and we arrange the expression we get
\be
q^\mu V_{1\mu}u=\left(\slashed{p}+\slashed{q}-M-\frac{1}{2}m^2\frac{\slashed{n}}{[n\cdot(p+q)]_{\bar{n}_2}}\right)-\left(\slashed{p}-M-\frac{1}{2}m^2\frac{\slashed{n}}{(n\cdot p)_{\bar{n}_1}}\right).
\ee
We recognize the inverse of the electron propagator and
\be
q^\mu V_{1\mu}[(p,\bar{n}_1),(p+q,\bar{n}_2)]=S_F^{-1}(p+q,\bar{n}_2)-S_F^{-1}(p,\bar{n}_1).
\ee

In the case where $m=0$, the terms with $\bar{n}$ associated vanish, and the standard case is recovered, the expression is the Ward-Takahashi identity. Now, in VSR, as we have new kind of vertex, we can proceed for the two external photon legs vertex in the same way, giving a $\bar{n}$ for each denominator with $n\cdot p$, so
\be
V_{2\mu\nu}[(p,\bar{n}_1),(p',\bar{n}_2),(p+q_1,\bar{n}_3),(p+q_2,\bar{n}_4)]=\frac{1}{2}m^2\slashed{n}\frac{n_\mu n_\nu}{(n\cdot p)_{\bar{n}_1}(n\cdot p')_{\bar{n}_2}}\left(\frac{1}{[n\cdot(p+q_1)]_{\bar{n}_3}}+\frac{1}{[n\cdot(p+q_2)]_{\bar{n}_4}}\right),
\ee
where $p'=p+q_1+q_2$. Again, we contract with an external photon leg, for instance $q_1$ and after performing a little bit of algebra we get
\be
q_1^\mu V_{2\mu\nu}[(p,\bar{n}_1),(p',\bar{n}_2),(p+q_1,\bar{n}_3),(p+q_2,\bar{n}_4)]=V_{1\nu}[(p,\bar{n}_1),(p+q_2,\bar{n}_4)]-V_{1\nu}[(p',\bar{n}_2),(p+q_1,\bar{n}_3)].
\ee
For the contraction with $q_2$ the result is
\be
q_2^\nu V_{2\mu\nu}[(p,\bar{n}_1),(p',\bar{n}_2),(p+q_1,\bar{n}_3),(p+q_2,\bar{n}_4)]=V_{1\mu}[(p,\bar{n}_1),(p+q_1,\bar{n}_3)]-V_{1\mu}[(p',\bar{n}_2),(p+q_2,\bar{n}_4)].
\ee

\begin{figure}[h!]
\centering
\includegraphics[scale=0.088]{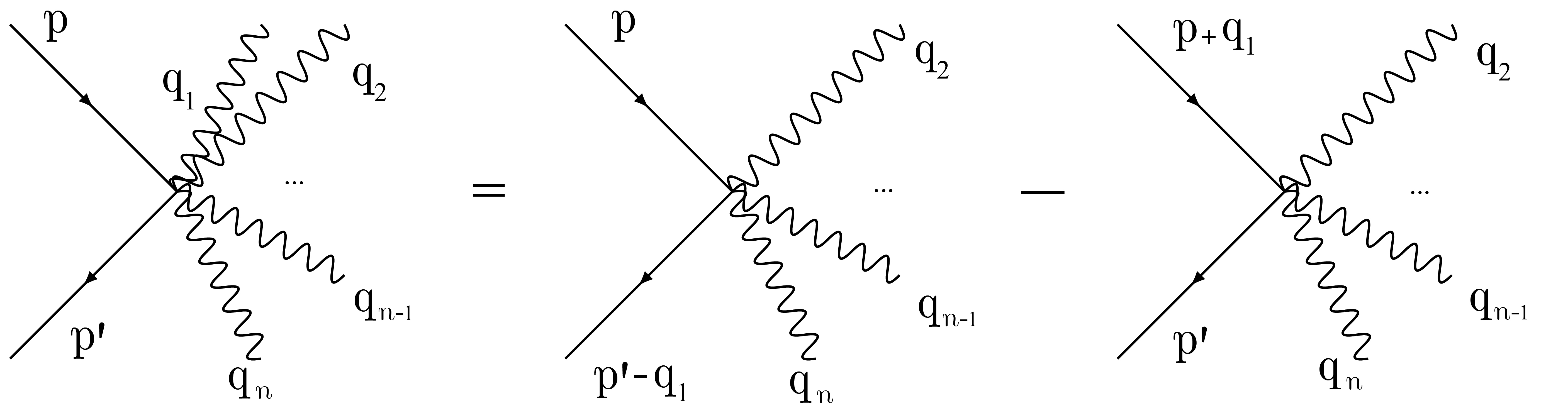}
\caption{Vertex with $n$ external photonic legs written as difference between two diagrams with one less leg.}
\label{fig:id}
\end{figure}

From this, we can write the contraction between a vertex with $n$ external photon legs with one external momentum of a photon as the difference between a vertex with $n-1$ external legs, without the leg whose momentum is the contracted, and another vertex with $n-1$ external legs, whose inner fermionic leg has momentum as the sum of the original inner fermionic momentum and the momentum contracted, as is shown in  figure \ref{fig:id}.

\section{Electron Self Energy}

\begin{figure}[h!]
\centering
\includegraphics[scale=0.4]{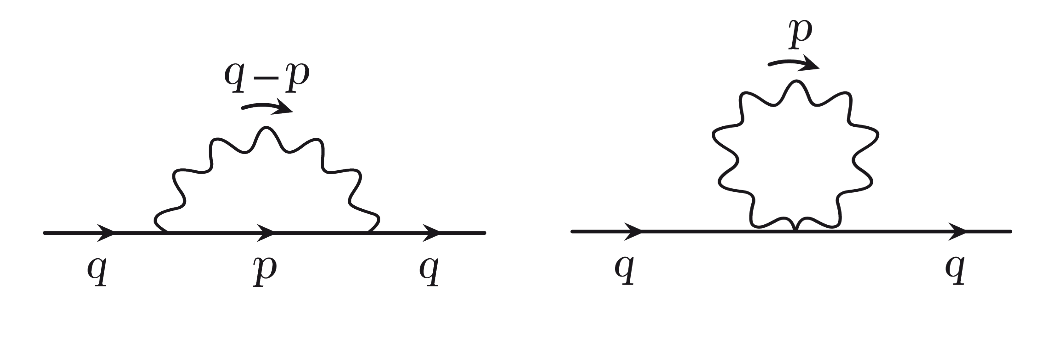}
\caption{Electron Self Energy diagrams in VSR.}
\label{fig:self}
\end{figure}

As the theory has a two external photon leg vertex we have another diagram, as is shown in the figure \ref{fig:self}. Nevertheless this new diagram does not contribute. We only have the following contribution from the first diagram given by
\be
\label{stot}
-i\Sigma(q)=-i\Sigma_{prev}(q)-i\Sigma_{new}(q),
\ee
where $-i\Sigma_{prev}$ is the contribution computed in \cite{Alfaro:2017umk} without photon mass, and $-i\Sigma_{new}$ is the new contribution coming from the terms with $m_\gamma$. The mathematical expressions to be computed are
\bea
\label{sprev}
-i\Sigma_{prev}(q)&=&(-i e)^2\int\frac{d^{2\omega}p}{(2\pi)^{2\omega}}\frac{1}{(p^2-M_e^2+i\varepsilon)((q-p)^2-m_\gamma^2+i\varepsilon)}\left(\gamma_\mu+\frac{1}{2}m^2\frac{\slashed{n}n_\mu}{n\cdot q n\cdot p}\right)\times \nn\\
& &\times\left(\slashed{p}+M-\frac{1}{2}m^2\frac{\slashed{n}}{n\cdot p}\right)\left(\gamma_\nu+\frac{1}{2}m^2\frac{\slashed{n}n_\nu}{n\cdot q n\cdot p}\right)g_{\mu\nu},
\eea
and
\bea
\label{snew}
-i\Sigma_{new}(q)&=&(-i e)^2m_\gamma^2\int\frac{d^{2\omega}p}{(2\pi)^{2\omega}}\frac{1}{(p^2-M_e^2+i\varepsilon)((q-p)^2-m_\gamma^2+i\varepsilon)}\left(\gamma_\mu+\frac{1}{2}m^2\frac{\slashed{n}n_\mu}{n\cdot q n\cdot p}\right)\times \nn\\
& &\times\left(\slashed{p}+M-\frac{1}{2}m^2\frac{\slashed{n}}{n\cdot p}\right)\left(\gamma_\nu+\frac{1}{2}m^2\frac{\slashed{n}n_\nu}{n\cdot q n\cdot p}\right)\left[\frac{n_\mu n_\nu}{[n\cdot(q-p)]^2}-\frac{(q_\mu-p_\mu)n_\nu+(q_\nu-p_\nu)n_\mu}{(q-p)^2[n\cdot(q-p)]}\right].
\eea
We use dimensional regularization and $2\omega$ is the dimension. Here we have omitted the subscripts $\bar{n}$ only to be more readable, but they are present. If we set $m=0$ and $m_\gamma=0$ we recover the standard computation. In addition, in equation (\ref{sprev}), $m_\gamma$ plays the role of the small photon mass introduced by hand in the standard case to regularize the infrared divergences, but here it is a parameter in the theory.\\

\begin{figure}[h!]
\centering
\includegraphics[scale=0.07]{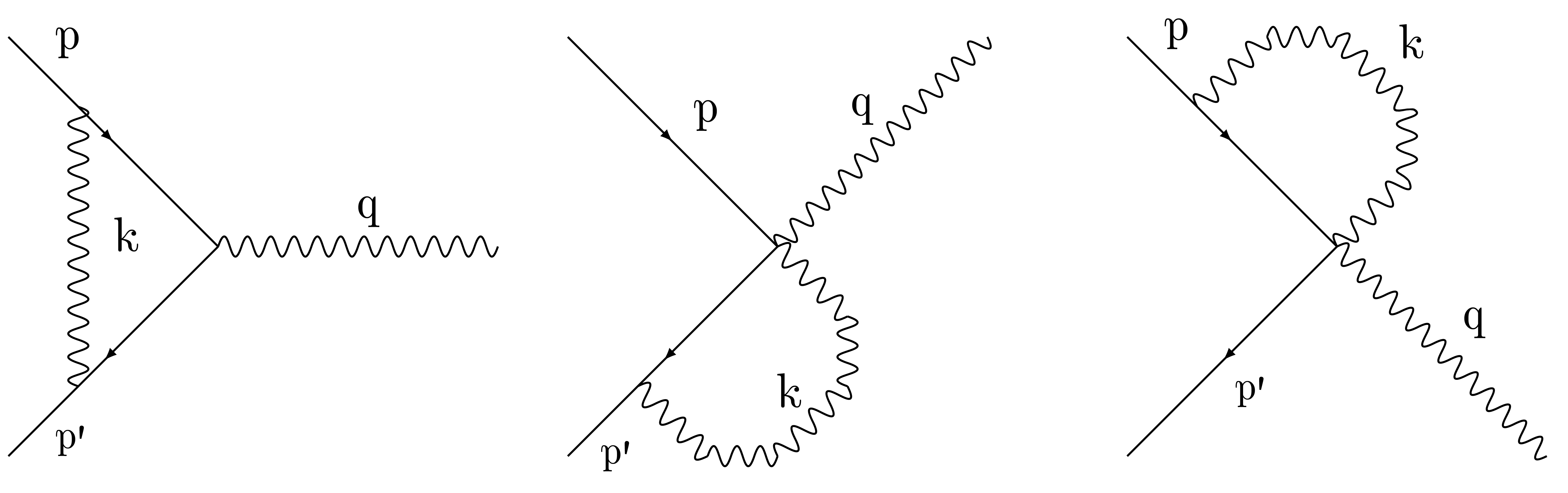}
\caption{Loop correction to the vertex.}
\label{fig:ward}
\end{figure}

We use the Ward identity to determine the relationship between the $\bar{n}$ for each term\cite{jauniverse}. In the figure \ref{fig:ward} we see the three diagrams for one loop correction to the vertex.\\

We write the diagrams in a mathematical way as
\bea
\bar{u}(p')\Lambda^\rho u(p)&=&\bar{u}(p',\bar{n}_1)V_1^\mu[(p'+k,\bar{n}_2),(p',\bar{n}_1)]S_F(p'+k,\bar{n}_2)V_1^\rho[(p'+k,\bar{n}_2),(p+k,\bar{n}_3)]S_F(p+k,\bar{n}_3)\times\nn\\
&&\times V_1^\nu[(p,\bar{n}_4),(p+k,\bar{n}_3)]\Delta_{\mu\nu}(k,\tilde{\tilde{n}}_1)u(p,\bar{n}_4)+\nn\\
&&+\bar{u}(p',\bar{\bar{n}}_1)V_1^\mu[(p'+k,\bar{\bar{n}}_2),(p',\bar{\bar{n}}_1)]S_F(p'+k,\bar{\bar{n}}_2)V_2^{\nu\rho}[(p,\bar{\bar{n}}_3),(p'+k,\bar{\bar{n}}_2),(p',\bar{\bar{n}}_1),(p+k,\bar{\bar{n}}_4)]\times\nn\\
&&\times\Delta_{\mu\nu}(k,\tilde{\tilde{n}}_2)u(p,\bar{\bar{n}}_3)+\nn\\
&&+\bar{u}(p',\tilde{n}_1)V_2^{\mu\rho}[(p-k,\tilde{n}_2),(p',\tilde{n}_1),(p,\tilde{n}_3),(p'-k,\tilde{n}_4)]S_F(p-k,\tilde{n}_2)V_1^\nu[(p,\tilde{n}_3),(p-k,\tilde{n}_2)]\times\nn\\
&&\times\Delta_{\mu\nu}(k,\tilde{\tilde{n}}_3)u(p,\tilde{n}_3)
\eea

We contract with the external leg $q$ and using the identities found in the last section, the first term gives
\bea
\bar{u}(p',\bar{n}_1)V_1^\mu[(p'+k,\bar{n}_2),(p',\bar{n}_1)]S_F(p+k,\bar{n}_3)V_1^\nu[(p,\bar{n}_4),(p+k,\bar{n}_3)]\Delta_{\mu\nu}(k,\tilde{\tilde{n}}_1)u(p,\bar{n}_4)\nn\\
-\bar{u}(p',\bar{n}_1)V_1^\mu[(p'+k,\bar{n}_2),(p',\bar{n}_1)]S_F(p'+k,\bar{n}_2)V_1^\nu[(p,\bar{n}_4),(p+k,\bar{n}_3)]\Delta_{\mu\nu}(k,\tilde{\tilde{n}}_1)u(p,\bar{n}_4),
\eea
the second
\bea
\bar{u}(p',\bar{\bar{n}}_1)V_1^\mu[(p'+k,\bar{\bar{n}}_2),(p',\bar{\bar{n}}_1)]S_F(p'+k,\bar{\bar{n}}_2)V_1^\nu[(p,\bar{\bar{n}}_3),(p+k,\bar{\bar{n}}_4)]\Delta_{\mu\nu}(k,\tilde{\tilde{n}}_2)u(p,\bar{\bar{n}}_3)\nn\\
-\bar{u}(p',\bar{\bar{n}}_1)V_1^\mu[(p'+k,\bar{\bar{n}}_2),(p',\bar{\bar{n}}_1)]S_F(p'+k,\bar{\bar{n}}_2)V_1^\nu[(p',\bar{\bar{n}}_1),(p'+k,\bar{\bar{n}}_2)]\Delta_{\mu\nu}(k,\tilde{\tilde{n}}_2)u(p,\bar{\bar{n}}_3),
\eea
and the latter after the change $k\to-k$
\bea
\bar{u}(p',\tilde{n}_1)V_1^\mu[(p+k,\tilde{n}_2),(p,\tilde{n}_3)]S_F(p+k,\tilde{n}_2)V_1^\nu[(p,\tilde{n}_3),(p+k,\tilde{n}_2)]\Delta_{\mu\nu}(k,\tilde{\tilde{n}}_3)u(p,\bar{n}_3)\nn\\
-\bar{u}(p',\tilde{n}_1)V_1^\mu[(p'+k,\tilde{n}_4),(p',\tilde{n}_1)]S_F(p+k,\tilde{n}_2)V_1^\nu[(p,\tilde{n}_3),(p+k,\tilde{n}_2)]\Delta_{\mu\nu}(k,\tilde{\tilde{n}}_3)u(p,\tilde{n}_3).
\eea

To satisfy the Ward Identity, we have
\bea
\bar{n}_1=\bar{\bar{n}}_1=\tilde{n}_1,\nn\\
\bar{n}_2=\bar{\bar{n}}_2=\tilde{n}_4,\nn\\
\bar{n}_3=\bar{\bar{n}}_4=\tilde{n}_2,\nn\\
\bar{n}_4=\bar{\bar{n}}_3=\tilde{n}_3,
\eea
and
\be
\label{nbarp}
\tilde{\tilde{n}}_1=\tilde{\tilde{n}}_2=\tilde{\tilde{n}}_3.
\ee

\begin{figure}[h!]
\centering
\includegraphics[scale=0.5]{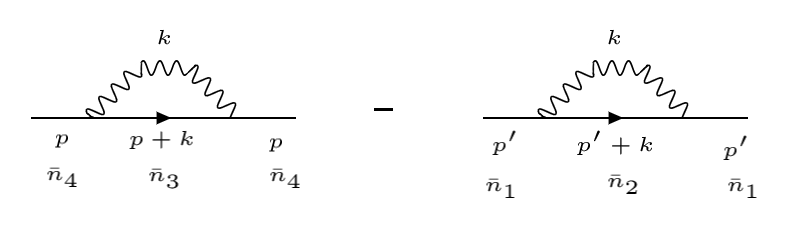}
\caption{Ward identity, vertex written as difference between two propagators with different electron momentum.}
\label{fig:warddiff}
\end{figure}

In this way, we have the scheme shown in the figure \ref{fig:warddiff}, defined by the expression
\bea
q_\rho \Lambda^\rho=V_1^\mu[(p+k,\bar{n}_3),(p,\bar{n}_4)]S_F(p+k,\bar{n}_3)V_1^\nu[(p,\bar{n}_4),(p+k,\bar{n}_3)]\Delta_{\mu\nu}(k,\tilde{\tilde{n}}_1)\nn\\
-V_1^\mu[(p'+k,\bar{n}_1),(p',\bar{n}_2)]S_F(p'+k,\bar{n}_2)V_1^\nu[(p',\bar{n}_1),(p'+k,\bar{n}_2)]\Delta_{\mu\nu}(k,\tilde{\tilde{n}}_1),
\eea
where we have omited the external legs $u$ and $\bar{u}$.\\

In the reference \cite{Alfaro:2017umk} the integrals with $\bar{n}$ are traded by a linear combination between the external legs and the null vector $n$ to restore a SIM(2) invariant expression, because the introduction of the Mandelstam-Leibbrandt prescription breaks the VSR invariance. It was shown the way to do it is
\be
\label{nbardef}
\bar{n}_\mu=-\frac{q^2}{2(n\cdot q)^2}n_\mu+\frac{q_\mu}{n\cdot q},
\ee
where $q$ is the external momentum in the diagram.\\

We observe in the figure \ref{fig:warddiff}, as $\bar{n}_3$ as $\bar{n}_4$ have the same external momentum $p$ and $\bar{n}_1$ with $\bar{n}_2$ share the external momentum $p'=p+k$, so $\bar{n}_1=\bar{n}_2$ and $\bar{n}_3=\bar{n}_4$. With this rule, we use the same $\bar{n}$ to compute the integrals in equation (\ref{sprev}) using the integration shown in \cite{Alfaro:2016pjw}, whose main integrals are listed in the appendix \ref{app-int}, then we trade the $\bar{n}$ with the linear combination in equation (\ref{nbardef})  and we have the same result that \cite{Alfaro:2017umk}.\\

On the other hand, because the equation (\ref{nbarp}) is satisfied, the two diagrams in the figure \ref{fig:warddiff} have the same $\bar{n}$ related to the photon propagator, but the external momenta are different. Here we cannot set $\bar{n}$ equal to zero for this case, because it does not respect the condition $n\cdot\bar{n}=1$. We need to use the equation (\ref{nbardef}) and select a common vector to both diagrams. The only one is the zero four vector and we proceed in the limit sense. Thus, we start with an arbitrary and common momentum $P$ and then we use the limit  $P\to0$ to eliminate the arbitrariness. Therefore, the Mandelstam-Leibbrandt prescription in the eq. (\ref{M-L}) after the replacement of $\bar{n}$ using the $P$ vector reads
\be
\label{mlnbar}
\frac{1}{n\cdot k}=\lim_{\eta\to0}\lim_{P\to0}\frac{-P^2n\cdot k+2P\cdot k n\cdot P}{(n\cdot k)(-P^2 n\cdot k+2P\cdot k n\cdot P)+i\eta},
\ee

where we have defined $\eta=2\epsilon(n\cdot P)^2$ which satisfies $\eta>0$. Notice that for $(n\cdot P)\neq 0$,  $lim_{\epsilon\to0}$ coincides with $lim_{\eta\to0}$, so we use as a definition the expression (\ref{mlnbar}). In the limit $P\to0$ and keeping $\eta$ not zero, the fraction $1/n\cdot k$ vanishes without problem. Using this rule, the final result of the self energy in equation (\ref{stot}) is
\be
-i\Sigma(q)=C\frac{\slashed{n}}{n\cdot q}+D\slashed{q}+E,
\ee
with
\bea
C&=&(-ie)^2m^2\left[-\frac{i}{16\pi^2}\int_0^1 dx\frac{1}{x}\log{\left(1+\frac{x^2q^2}{(1-x)M_e^2-xq^2+xm_\gamma^2-i\varepsilon}\right)}\right.\nn\\
&&\left.+\frac{2i}{(4\pi)^{\omega}}\int_0^1 dx\frac{\Gamma(2-\omega)}{[(1-x)M_e^2-x(1-x)q^2+xm_\gamma^2-i\varepsilon]^{2-\omega}}+\frac{i}{8\pi^2}\int_0^1 dx \log{\left(1+\frac{m_\gamma^2(1-x)}{xM_e^2-x(1-x)q^2}\right)}\right],\\
D&=&-2(-ie)^2(\omega-1)\frac{i}{(4\pi)^{\omega}}\int_0^1 dx\frac{x\Gamma(2-\omega)}{[(1-x)M_e^2-x(1-x)q^2+xm_\gamma^2-i\varepsilon]^{2-\omega}}\nn\\
&&+\frac{i}{8\pi^2}\int_0^1 dx \log{\left(1+\frac{m_\gamma^2(1-x)}{xM_e^2-x(1-x)q^2}\right)},\\
E&=&(-ie)^2 2\omega M\frac{i}{(4\pi)^{\omega}}\int_0^1 dx\frac{\Gamma(2-\omega)}{[(1-x)M_e^2-x(1-x)q^2+xm_\gamma^2-i\varepsilon]^{2-\omega}}\nn\\
&&+M\frac{i}{8\pi^2}\int_0^1 dx \log{\left(1+\frac{m_\gamma^2(1-x)}{xM_e^2-x(1-x)q^2}\right)}.
\eea

We notice the self energy is not infrared divergent, because the photon mass regularize the possible divergence. Moreover, the photon mass is not an ad hoc addition but it is a theory parameter. 

\section{Coulomb Scattering}

Here we will review the electron scattering due an external static electric field. At tree level we have
\be
i\M=\bar{u}(p')\left[(-ie)\left(\gamma^0+\frac{1}{2}m^2\frac{\slashed{n}n^0}{n\cdot p n\cdot p'}\right)A_0(q)\right]u(p),
\ee
where the only non zero component in $A_\mu$ is $A_0=\frac{Ze^2}{\vert \vec{q}\vert^2}$. Therefore,
\bea
\label{msq}
\vert\M \vert^2&=&\frac{2Ze^4}{\vert \vec{q}\vert^4}\left[p^0p'^0+\vec{p}\cdot\vec{p'}+M_e^2-m^2+\frac{1}{2}m^2\left(\frac{n\cdot p}{n\cdot p'}+\frac{n\cdot p'}{n\cdot p}\right)+m^2n^0(p^0-p'^0)\left(\frac{1}{n\cdot p}-\frac{1}{n\cdot p'}\right)\right.\nn\\
&&\left.+m^2(M_e^2-p\cdot p')\frac{(n^0)^2}{n\cdot p n\cdot p'}\right].
\eea

Since the external field only changes the direction of the momentum, but it does not change its magnitude, we have $\vert \vec{p}\vert=\vert\vec{p'}\vert$. Then, as $\vec{p'}=\vec{p}+\vec{q}$ we get
\be
\vert \vec{q}\vert^2=4\vert\vec{p}\vert^2\sin^2\frac{\theta}{2},
\ee
where $\theta$ is the deflection angle. Moreover, since the energy is conserved, $p^0=p'^0=E$ and $p^2=M_e^2$, then the equation (\ref{msq}) is now
\be
\vert\M \vert^2=\frac{Ze^4}{8\vert\vec{p}\vert^4\sin^4\frac{\theta}{2}}\left[2E^2-2\vert\vec{p}\vert^2\sin^2\frac{\theta}{2}-m^2+\frac{1}{2}m^2\left(\frac{n\cdot p}{n\cdot p'}+\frac{n\cdot p'}{n\cdot p}\right)-m^2\vert\vec{p}\vert^2\sin^2\frac{\theta}{2}\frac{(n^0)^2}{n\cdot p n\cdot p'}\right].
\ee
We use the frame of reference where $n=(1,0,0,1)$, therefore
\bea
\vert\M \vert^2&=&\frac{Ze^4}{8\vert\vec{p}\vert^4\sin^4\left(\frac{\theta}{2}\right)}\left[2E^2-2\vert\vec{p}\vert^2\sin^2\left(\frac{\theta}{2}\right)-m^2+\frac{1}{2}m^2\left(\frac{E-\vert\vec{p}\vert\sin\eta\sin\phi}{E-\vert\vec{p}\vert\sin\eta\sin(\phi-\theta)}+\frac{E-\vert\vec{p}\vert\sin\eta\sin(\phi-\theta)}{E-\vert\vec{p}\vert\sin\eta\sin\phi}\right)\right.\nn\\
&&\left.-m^2\vert\vec{p}\vert^2\sin^2\left(\frac{\theta}{2}\right)\frac{1}{(E-\vert\vec{p}\vert\sin\eta\sin\phi)(E-\vert\vec{p}\vert\sin\eta\sin(\phi-\theta))}\right],
\eea
where $\eta$ is the angle between $\vec{n}$ and the normal vector to the plane where the scattering occurs. In addition, $\phi$ is the angle between the projection of $\vec{n}$ in the plane where the scattering takes place and $-\vec{p}$. With this, the cross section is given by
\bea
\frac{d\sigma}{d\Omega}&=&\frac{Z\alpha^2}{4\vert\vec{p}\vert^2\beta^2\sin^4\left(\frac{\theta}{2}\right)}\left[1-\beta^2\sin^2\left(\frac{\theta}{2}\right)-\frac{m^2}{2M_e^2}+\frac{m^2}{4M_e^2}\left(\frac{1-\beta\sin\eta\sin\phi}{1-\beta\sin\eta\sin(\phi-\theta)}+\frac{1-\beta\sin\eta\sin(\phi-\theta)}{1-\beta\sin\eta\sin\phi}\right)\right.\nn\\
&&\left.-\frac{m^2}{2M_e^2}\beta^2\sin^2\left(\frac{\theta}{2}\right)\frac{1}{(1-\beta\sin\eta\sin\phi)(1-\beta\sin\eta\sin(\phi-\theta))}\right].
\eea
Here we have got a VSR Mott formula, where if set $m=0$ we recover the classic result. 

\section{Infrared Divergences Cancellation}
Now, we move on to the radiative corrections to the tree level Coulomb Scattering. In the standard case is well known that the infrared divergences in the one loop corrections, as in the vertex and the fermionic legs, are cancelled by the bremsstrahlung process, with the soft photon emission. So, here we will compute these corrections, under the assumption that the radiated  photon momentum $k$  is small. Therefore
\be
\vert\M_{brem}\vert^2=\int\frac{d^3\vec{k}}{(2\pi)^22\sqrt{\vert\vec{k}\vert^2+m_\gamma^2}}\sum_{\lambda=1,2} e^2\left[\frac{p'^\mu\varepsilon_\mu(k)}{p'\cdot k}-\frac{p^\mu\varepsilon_\mu(k)}{p\cdot k}\right]\left[\frac{p'^\nu\varepsilon^*_\nu(k)}{p'\cdot k}-\frac{p^\nu\varepsilon^*_\nu(k)}{p\cdot k}\right]\vert\M \vert^2,
\ee
where the sum is over the two photon polarizations and $\vert\M\vert^2$ is the tree level quantity, given by the equation (\ref{msq}). The sum over the polarizations in VSR is given by
\be
\label{sumpol}
\sum_{\lambda=1}^2\varepsilon_\mu(k)\varepsilon^*_\nu(k)=-g_{\mu\nu}-\frac{m_\gamma^2}{(n\cdot k)^2}n_\mu n_\nu+\frac{1}{n\cdot k}(k_\mu n_\nu+k_\nu n_\mu).
\ee
The full computation of this sum is shown in the appendix \ref{sum-pol}. The algebra is straightforward, then,  
\bea
\label{bremm}
\vert\M_{brem}\vert^2&=&e^2\int\frac{d^3\vec{k}}{(2\pi)^22\sqrt{\vert\vec{k}\vert^2+m_\gamma^2}}\left[\frac{2}{p\cdot k p'\cdot k}\left(p\cdot p'+m_\gamma^2\frac{n\cdot p n\cdot p'}{(n\cdot k)^2}\right)-\frac{1}{(p'\cdot k)^2}\left(M_e^2+m_\gamma^2\frac{(n\cdot p')^2}{(n\cdot k)^2}\right)\right.\nn\\
&&\left.-\frac{1}{(p\cdot k)^2}\left(M_e^2+m_\gamma^2\frac{(n\cdot p)^2}{(n\cdot k)^2}\right)\right]\vert\M \vert^2.
\eea

The vertex correction is easy to compute. Again, we work in the small $k$ approximation. To get the contributions to the $\M$ matrix at the order that we are working, we need the interference with the diagram at tree level. After a long computation we have got
\be
\label{vertex}
\vert\M_{mix}\vert^2=-2e^2\int\frac{d^3\vec{k}}{(2\pi)^22\sqrt{\vert\vec{k}\vert^2+m_\gamma^2}}\frac{1}{p\cdot k p'\cdot k}\left(p\cdot p'+m_\gamma^2\frac{n\cdot p n\cdot p'}{(n\cdot k)^2}\right)\vert\M \vert^2.
\ee
We can observe the above expression cancel the first term in equation (\ref{bremm}). The external legs corrections cancel the latter terms. We see that the bremmstrahlung cancel the radiative corrections at one loop as in the standard QED. As consequence, it is impossible to get a signal of the photon mass in the process to distinguish between the standard case and VSR, because the behavior is the same. Here, the photon mass is absent as usual, so, there is not difference with the standard case at this order.\\ 

We notice that we can fix the light cone gauge (lcg), $n\cdot A=0$. Using this gauge we see that all vertices with more than one
photon leg do not contribute to the matrix elements. In Appendix C we have obtained the lcg photon propagator $\Delta_{\mu\nu}$ and checked that
it satisfies $\Delta_{\mu\nu} n^\mu=0$. This property of the propagator guarantees that only the one photon vertex will contribute
to any process. This may pave the way to show that the perturbative expansion in VSR QED has the same validity as in QED, but to prove this will need extra work.\\

Additionally, our calculation shows that there is an exact cancelation of the infrared divergences if we add the contribution of soft photons.
That is a  Bloch-Nordsieck \cite{bn,yennie,chung,kulish} treatment may be possible to implement in VSR QED. The proof at all orders is a very interesting issue. However, this idea is out of the scope of our manuscript. It is an important question that should be addressed in a future work.

\section{Conclusions}

We have reviewed the possibility in VSR to have a  gauge invariant photon mass. Although the current experimental bounds tell us it should be too small, it is a new feature of the model to distinguish between the standard case and VSR. Despite $m_\gamma$ is a free parameter that we can set equal to zero without problems, we have explored the consequences to consider a non null value.\\

A modified Maxwell equations was presented. As the new contributions from the VSR sector are multiplied by the photon mass, which is very small, a classical experiment to measure departures from the standard results should not be useful. Nevertheless the equations are presented in any case.\\

We have computed the electron self energy considering the photon mass and we have seen it does not present infrared divergencies. Moreover, in the loop corrections there is not signal of this mass in the observables. We have computed the Coulomb scattering, and the photon mass is cancelled as in the standard case at one loop due to the radiative corrections in the low energy region.\\

In the cross section calculation we observe a small difference with the standard result, that it should be considered to measure. Although the photon mass is absent, it could be found a signal of $n$ in this observable. We expect it is small due it is proportional to $m^2/M_e^2$.\\

\begin{acknowledgements}

We thank the participants of the conference ``La parte y el todo'' in Afunalhue, Chile for useful comments and discussions. The work of A.Soto is supported by the CONICYT-PFCHA/Doctorado Nacional/2017-21171194 and
Fondecyt 1150390. The work of J. Alfaro is partially supported by Fondecyt 1150390 and CONICYT-PIA-ACT1417.\\

\end{acknowledgements}
\vspace{-20pt}

\begin{appendix}

\section{Sum over polarizations} \label{sum-pol}
We write the sum of the polarizations as
\be
\sum_{\lambda=1}^2\varepsilon_\mu(k)\varepsilon^*_\nu(k)=Ag_{\mu\nu}+Bk_\mu k_\nu+Cn_\mu n_\nu+D(k_\mu n_\nu+k_\nu n_\mu).
\ee
We use the following conditions to get the coefficients
\bea
k^\mu \varepsilon_\mu=0, \label{cond1}\\
n^\mu \varepsilon_\mu=0, \label{cond2}\\
g^{\mu\nu}\varepsilon^\lambda_\mu\varepsilon^{\lambda*}_\nu=-1. \label{cond3}
\eea

From equations (\ref{cond1}), (\ref{cond2}) and  (\ref{cond3}) we get
\bea
0=(A+Bk^2+Dn\cdot k)k_\nu+(Cn\cdot k+Dk^2)n_\nu, \label{cond1b}\\
0=Bn\cdot k k_\nu+(A+Dn\cdot k)n_\nu, \label{cond2b}\\
-2=4A+Bk^2+2Dn\cdot k, \label{cond3b}
\eea
respectively. From equation (\ref{cond2b}) we get $B=0$ and $A=-Dn\cdot k$. We insert it in equation (\ref{cond1b}), from there we have a consistency relation in the first parentheses and $C=-D\frac{k^2}{n\cdot k}$. And we plug $A$ and $B$ in the equation (\ref{cond3b}) to get $D$. Finally
\bea
A=-1,\\
B=0,\\
C=-\frac{k^2}{(n\cdot k)^2},\\
D=\frac{1}{n\cdot k}.
\eea
Thus, we use $k^2=m_\gamma^2$ in $C$ and the sum of the polarizations is
\be
\sum_{\lambda=1}^2\varepsilon_\mu(k)\varepsilon^*_\nu(k)=-g_{\mu\nu}-\frac{m_\gamma^2}{(n\cdot k)^2}n_\mu n_\nu+\frac{1}{n\cdot k}(k_\mu n_\nu+k_\nu n_\mu).
\ee

\section{Integration with $(n\cdot p)^{-1}$} \label{app-int}
Here we quote the main integrals needed to compute our expressions from reference \cite{Alfaro:2016pjw}. They are
\be
\label{ap1}
\int dp\frac{1}{(p^2+2p\cdot q-m^2)^a}\frac{1}{(n\cdot p)^b}=(-1)^{a+b}i\pi^\omega(-2)^b\frac{\Gamma(a+b)}{\Gamma(a)\Gamma(b)}(\bar{n}\cdot q)^b\int_0^1 dt t^{b-1}\frac{1}{(m^2+q^2-2(n\cdot q)(\bar{n}\cdot q)t)^{a+b-\omega}},
\ee
with $\omega=d/2$.\\

Taking a derivative with respect to $q$ in (\ref{ap1}) we get
\bea
\int d^d p \frac{p_{\mu}}{(p^2 + 2 p \cdot q - m^2)^{a + 1}} \frac{1}{(n
 \cdot p)^b} &=&
(- 1)^{a + b} i \pi^{\omega} (- 2)^{b - 1} \frac{\Gamma (a + b -
\omega)}{\Gamma (a + 1) \Gamma (b)} (\bar{n} \cdot q)^{b - 1} b
\bar{n}_{\mu} \int^1_0 d t t^{b - 1} \frac{1}{[m^2 + q^2 - 2 (n  \cdot q)
(\bar{n} \cdot q) t]^{a + b - \omega}}\nn\\
&+&(- 1)^{a + b} i \pi^{\omega} (- 2)^b \frac{\Gamma (a + b + 1 -
\omega)}{\Gamma (a + 1) \Gamma (b)} (\bar{n}  \cdot q)^b \int^1_0 d t t^{b -
1} \frac{q_{\mu} - t (n  \cdot q \bar{n}_{\mu} + \bar{n}  \cdot q
n_{\mu})}{[m^2 + q^2 - 2 (n  \cdot q) (\bar{n}  \cdot q) t]^{a + b + 1 -
\omega}}.\nn\\
\eea

With a second derivative in (\ref{ap1}): 

\bea
\int d^d p \frac{p_{\mu} p_{\nu}}{(p^2 + 2 p \cdot q - m^2)^{a + 2}}
\frac{1}{(n \cdot p)^b} &=&(- 1)^{a + b} i \pi^{\omega} (- 2)^{b - 2}\times\nn\\
&&\times\left\{ \frac{\Gamma (a + b -
\omega)}{\Gamma (a + 2) \Gamma (b - 1)} (\bar{n} \cdot q)^{b - 2} b
\bar{n}_{\mu} \bar{n}_{\nu} \int^1_0 d t t^{b - 1} \frac{1}{(m^2 + q^2 - 2 (n
\cdot q) (\bar{n} \cdot q) t)^{a + b - \omega}}\right.\nn\\
&&\left.- 2 \frac{\Gamma (a + b +1 - \omega)}{\Gamma (a + 2) \Gamma (b)} (\bar{n} \cdot q)^{b - 1} b
\bar{n}_{\mu} \int^1_0 d t t^{b - 1} \frac{q_{\nu} - t (n \cdot q
\bar{n}_{\nu} + \bar{n} \cdot q n_{\nu})}{(m^2 + q^2 - 2 (n \cdot q)
(\bar{n} \cdot q) t)^{a + b + 1 - \omega}}\right.\nn\\
&&\left. - 2 \frac{\Gamma (a + b + 1 -
\omega)}{\Gamma (a + 2) \Gamma (b)} \left( \bar{n} \cdot q \right)^{b - 1} b
\bar{n}_{\nu} \int^1_0 d t t^{b - 1} \frac{q_{\mu} - t (n \cdot q
\bar{n}_{\mu} + \bar{n} \cdot q n_{\mu})}{(m^2 + q^2 - 2 (n \cdot q)
(\bar{n} \cdot q) t)^{a + b + 1 - \omega}} \right.\nn\\
&&\left.+ 4 \frac{\Gamma (a + b + 2 -
\omega)}{\Gamma (a + 2) \Gamma (b)} (\bar{n} \cdot q)^b \int^1_0 d t t^{b -
1} \frac{[q_{\nu} - t (n \cdot q \bar{n}_{\nu} + \bar{n} \cdot q n_{\nu})]
[q_{\mu} - t (n \cdot q \bar{n}_{\mu} + \bar{n} \cdot q n_{\mu})]}{(m^2 +
q^2 - 2 (n \cdot q) (\bar{n} \cdot q) t)^{a + b + 2 - \omega}} \right.\nn\\
&&\left.- 2
\frac{\Gamma (a + b + 1 - \omega)}{\Gamma (a + 2) \Gamma (b)} (\bar{n} \cdot
q)^b \int^1_0 d t t^{b - 1} \frac{g_{\mu \nu} - t (n_{\nu} \bar{n}_{\mu} +
\bar{n}_{\nu} n_{\mu})}{(m^2 + q^2 - 2 (n \cdot q) (\bar{n} \cdot q) t)^{a
+ b + 1 - \omega}} \right\}
\eea


\section{Light cone gauge (lcg)}

We can fix the lcg:
\[ n\cdot A=0 \]
In lcg, we get:
\[ n \cdot D=n \cdot \partial +i e n \cdot A=n \cdot \partial \]
Therefore in lcg we have the standard QED vertex only, therefore the
perturbative series has the same validity as in QED.\\

To be more
explicit, we compute the lcg propagator with a photon mass:
\begin{equation}
{\cal L}_{gauge+gf} =- \frac{1}{4} F_{\mu \nu} F^{\mu \nu} - \frac{1}{2}
m_{\gamma}^{2}  (n^{\alpha} F_{\mu \alpha} ) \frac{1}{(n \cdot \partial
	)^{2}}  (n_{\beta} F^{\mu \beta} ) - \frac{1}{2 \xi}  (n_{\mu} A^{\mu} )^{2}
.
\end{equation}
$\xi\to0$ limit must be taken.\\

Performing some algebraic manipulations we get
\begin{equation}
{\cal L}_{gauge+gf} = \frac{1}{2} A_{\nu}  \left[ ( \partial^{2} +m_{\gamma}^{2}
)g^{\mu \nu} - \partial^{\mu} \partial^{\nu} +m_{\gamma}^{2}  \frac{n^{\mu}
	n^{\nu} \partial^{2}}{(n \cdot \partial )^{2}} -m_{\gamma}^{2} 
\frac{\partial^{\mu} n^{\nu} + \partial^{\nu} n^{\mu}}{n \cdot \partial} -
\frac{1}{\xi} n^{\mu} n^{\nu} \right] A_{\mu} .
\end{equation}
That is, after performing $\xi\to0$:
\begin{equation}
\label{propph} \Delta_{\mu \nu} (p, \bar{n} ) =- \frac{i}{p^{2}
	-m_{\gamma}^{2}}  \left[ g_{\mu \nu} + \frac{m_{\gamma}^{2}}{(n \cdot
	p)_{\bar{n}}^{2}} n_{\mu} n_{\nu} - \frac{1}{(n \cdot p)_{\bar{n}}} (p_{\mu}
n_{\nu} +p_{\nu} n_{\mu} ) \right] ,
\end{equation}

It satisfies:
\[ \Delta_{\mu \nu} (p, \bar{n} )  n^{\nu} =0 \]
It is clear that this propagator contracted with one vertex containing more
that one photon leg will give zero, because all those vertices are
proportional to tensor powers of $n_{\mu}$.
\end{appendix}


\begin{thebibliography}{13}

\bibitem{Cohen:2006ky} 
  A.~G.~Cohen and S.~L.~Glashow,
  ``Very special relativity,''
  Phys.\ Rev.\ Lett.\  {\bf 97}, 021601 (2006)
  [hep-ph/0601236].

\bibitem{Cohen:2006ir} 
  A.~G.~Cohen and S.~L.~Glashow,
  ``A Lorentz-Violating Origin of Neutrino Mass?,''
  hep-ph/0605036.
  
  
\bibitem{Peskin:1995ev} 
  M.~E.~Peskin and D.~V.~Schroeder,
  ``An Introduction to quantum field theory,"  Addison-Wesley (1995), chapter 7.5.
  
\bibitem{Cheon:2009zx} 
  S.~Cheon, C.~Lee and S.~J.~Lee,
  ``SIM(2)-invariant Modifications of Electrodynamic Theory,''
  Phys.\ Lett.\ B {\bf 679}, 73 (2009)
  [arXiv:0904.2065 [hep-th]].
  
\bibitem{Alfaro:2015fha} 
  J.~Alfaro, P.~Gonzalez and R.~Avila,
  ``Electroweak standard model with very special relativity,''
  Phys.\ Rev.\ D {\bf 91}, 105007 (2015)
  Addendum: [Phys.\ Rev.\ D {\bf 91}, no. 12, 129904 (2015)]
  [arXiv:1504.04222 [hep-ph]].

\bibitem{Alfaro:2013uva} 
  J.~Alfaro and V.~O.~Rivelles,
  ``Non Abelian Fields in Very Special Relativity,''
  Phys.\ Rev.\ D {\bf 88}, 085023 (2013)
  [arXiv:1305.1577 [hep-th]].
 
\bibitem{Bonetti:2016cpo} 
  L.~Bonetti, J.~Ellis, N.~E.~Mavromatos, A.~S.~Sakharov, E.~K.~G.~Sarkisyan-Grinbaum and A.~D.~A.~M.~Spallicci,
  ``Photon Mass Limits from Fast Radio Bursts,''
  Phys.\ Lett.\ B {\bf 757}, 548 (2016)
  [arXiv:1602.09135 [astro-ph.HE]].
  
\bibitem{Tanabashi:2018oca} 
  M.~Tanabashi {\it et al.} [Particle Data Group],
  ``Review of Particle Physics,''
  Phys.\ Rev.\ D {\bf 98}, no. 3, 030001 (2018).


\bibitem{Mandelstam:1982cb} 
  S.~Mandelstam,
  ``Light Cone Superspace and the Ultraviolet Finiteness of the N=4 Model,''
  Nucl.\ Phys.\ B {\bf 213}, 149 (1983).
  
\bibitem{Leibbrandt:1983pj} 
  G.~Leibbrandt,
  ``The Light Cone Gauge in Yang-Mills Theory,''
  Phys.\ Rev.\ D {\bf 29}, 1699 (1984).
  
\bibitem{Alfaro:2016pjw} 
  J.~Alfaro,
  ``Mandelstam-Leibbrandt prescription,''
  Phys.\ Rev.\ D {\bf 93}, no. 6, 065033 (2016)
  Erratum: [Phys.\ Rev.\ D {\bf 94}, no. 4, 049901 (2016)]
  [arXiv:1603.06453 [hep-th]].
  
\bibitem{Alfaro:2017umk} 
  J.~Alfaro,
  ``A $Sim(2)$ invariant dimensional regularization,''
  Phys.\ Lett.\ B {\bf 772}, 100 (2017)
  [arXiv:1704.02299 [hep-th]].
\bibitem{jauniverse}  J. Alfaro,  ``Feynman Rules, Ward Identities and Loop Corrections in Very Special Relativity Standard Model,'' Universe 2019, 5(1), 16; 
      
\bibitem{bn} F. Bloch and A. Nordsieck, Phys. Rev., 52, 54 (1937).
\bibitem{yennie}D. Yennie, S. Frautschi, and H. Suura, Ann: Phys. (New York), 13, 379 (1961).
\bibitem{chung}V. Chung, Phys. Rev.0140B, 1110 (1965).
\bibitem{kulish} P. P. Kulish and L. D. Faddeev, Theor. Math. Phys.4, 745 (1970) [Teor. Mat. Fiz.4, 153(1970)].
\end{thebibliography}
\end{document}